\documentstyle[12pt]{article}
\begin{document}
\thispagestyle{empty}

\def\ve#1{\mid #1\rangle}
\def\vc#1{\langle #1\mid}

\newcommand{\p}[1]{(\ref{#1})}
\newcommand{\be}{\begin{equation}}
\newcommand{\ee}{\end{equation}}
\newcommand{\sect}[1]{\setcounter{equation}{0}\section{#1}}

\newcommand{\vs}[1]{\rule[- #1 mm]{0mm}{#1 mm}}
\newcommand{\hs}[1]{\hspace{#1mm}}
\newcommand{\mb}[1]{\hs{5}\mbox{#1}\hs{5}}
\newcommand{\Db}{{\overline D}}
\newcommand{\bea}{\begin{eqnarray}}
\newcommand{\eea}{\end{eqnarray}}
\newcommand{\wt}[1]{\widetilde{#1}}
\newcommand{\und}[1]{\underline{#1}}
\newcommand{\ov}[1]{\overline{#1}}
\newcommand{\sm}[2]{\frac{\mbox{\footnotesize #1}\vs{-2}}
           {\vs{-2}\mbox{\footnotesize #2}}}
\newcommand{\prt}{\partial}
\newcommand{\eps}{\epsilon}

\newcommand{\R}{\mbox{\rule{0.2mm}{2.8mm}\hspace{-1.5mm} R}}
\newcommand{\Z}{Z\hspace{-2mm}Z}

\newcommand{\cd}{{\cal D}}
\newcommand{\cg}{{\cal G}}
\newcommand{\ck}{{\cal K}}
\newcommand{\cw}{{\cal W}}

\newcommand{\vj}{\vec{J}}
\newcommand{\vl}{\vec{\lambda}}
\newcommand{\vz}{\vec{\sigma}}
\newcommand{\vt}{\vec{\tau}}
\newcommand{\vw}{\vec{W}}
\newcommand{\poiss}{\stackrel{\otimes}{,}}

\def\l#1#2{\raisebox{.2ex}{$\displaystyle
  \mathop{#1}^{{\scriptstyle #2}\rightarrow}$}}
\def\r#1#2{\raisebox{.2ex}{$\displaystyle
 \mathop{#1}^{\leftarrow {\scriptstyle #2}}$}}

\renewcommand{\thefootnote}{\fnsymbol{footnote}}
\newpage
\setcounter{page}{0}
\pagestyle{empty}
\begin{flushright}
{March 1998}\\
{IIMAS-UNAM-80}\\
{JINR E2-98-49}\\
{solv-int/9803010}
\end{flushright}
\vs{8}
\begin{center}
{\LARGE {\bf The solution of the $N=(0|2)$}}\\[0.6cm]
{\LARGE {\bf superconformal f--Toda lattice}}\\[1cm]

\vs{8}

{\large V.B. Derjagin$^{a,1}$, A.N. Leznov$^{a,b,c,2}$ and
A. Sorin$^{c,3}$} {}~\\ \quad \\

{\em {~$~^{(a)}$ Institute for High Energy Physics,}}\\
{\em 142284 Protvino, Moscow Region, Russia}\\

\vspace{2mm}

{\em  {~$~^{(b)}$} Instituto de Investigaciones en Matem\'aticas
  Aplicadas y en Sistemas}~\\
{\em Universidad Nacional Aut\`onoma de M\`exico} \\
{\em Apartado Postal 48-3, 62251 Cuernavaca,    \\
     Morelos, M\'exico} \\

\vspace{2mm}

{\em {~$~^{(c)}$} Bogoliubov Laboratory of Theoretical Physics, JINR,}\\
{\em 141980 Dubna, Moscow Region, Russia}\\

\end{center}
\vs{8}

\centerline{ {\bf Abstract}}
\vs{4}

The general solution of the two--dimensional integrable generalization of the
f--Toda chain with fixed ends is explicitly presented in
terms of matrix elements of various fundamental representations of the
$SL(n|n-1)$ supergroup. The dominant role of the representation theory of
graded Lie algebras in the problem of constructing integrable mappings and
lattices is demonstrated.

\vfill

{\em E-Mail:\\
1) derjagin@mx.ihep.su\\
2) leznov@ce.ifisicam.unam.mx\\
3) sorin@thsun1.jinr.dubna.su }
\newpage
\pagestyle{plain}
\renewcommand{\thefootnote}{\arabic{footnote}}
\setcounter{footnote}{0}

\section{Introduction}

In paper \cite{dls}, the one--dimensional ($1D$) f-Toda
mapping (chain) responsible for the existence of the $N=2$ supersymmetric
Nonlinear Schr\"odinger (NLS) hierarchy \cite{rk} was introduced. Using this
mapping as a starting point, the Hamiltonian structure and recursion operator
connecting all evolution systems of the $N=2$ NLS hierarchy were explicitly
constructed there. On the other hand, it has been proved in \cite{ls2}
that the $1D$ f-Toda chain with fixed ends is exactly integrable system,
and the method of constructing its general solutions in terms of a
perturbative series with a finite number of terms was proposed. However,
these calculations are sufficiently boring, and hence the explicit
solution was presented only for a very particular case \cite{ls2}.
Interest in the general solution of the $1D$ f--Toda chain with fixed ends
is mainly motivated by the fact that they, in turn, allow one to construct
explicitly multi--soliton solutions of equations belonging the $N=2$ NLS
hierarchy.

After this, the $N=2$ supersymmetric Toda lattice hierarchy whose first
bosonic flow is equivalent to the $1D$ f-Toda chain was constructed, and its
Hamiltonian and Lax--pair descriptions were developed in \cite{bs}.

The aim of the present letter is to generalize the $1D$ f--Toda chain
to the two-dimensional case and to construct its general solutions.
It turns out that the integrable generalization --- $2D$ f--Toda lattice
--- actually exists and its general solutions in the case of fixed ends
can be expressed in terms of
matrix elements of different fundamental representations of the
$SL(n|n-1)$ supergroup. Our construction is mainly based on results of Ref.
\cite{l}, where the method of constructing a wide class of integrable systems
related to the $sl(n)$ algebras has been proposed. We observe that in a
framework of this method there exists some hidden, omitted in \cite{l}
non--trivial possibility for deriving a new class of integrable systems. The
developed here method is applied to both the case of algebras $sl(n)$
and superalgebras $sl(n|n-1)$. The $2D$ f--Toda lattice just belongs to
a new class and gives the first, simplest representative that possesses
the $N=(0|2)$ superconformal symmetry.

\section{Structure of $sl(n|n-1)$ and $sl(2n-1)$ (super)algebras}
In this section we shortly summarize the main facts concerning the
(super)algebras $sl(n|n-1)$ and $sl(2n-1)$ which we use in what follows
(for more detail, see \cite{l1,ls0,fss} and references therein).
We consider them simultaneously keeping the explicit dependence of all
relevant formulae on their Grassmann nature via the factor ${\cal P}$
which is ${\cal P}=0$ or ${\cal P}=1$, respectively.

The (super)algebra $sl(2n-1)$ ($sl(n|n-1)$) can be generated by a set of
$4n(n-1)$ graded $(2n-1)\times (2n-1)$ matrices with zero (super)trace.
In the Serre-Chevalley basis its defining commutation relations
are\footnote{Hereafter, we understand that there is no
summation over repeated indices.}
\begin{eqnarray}
[H_i , H_j]=0, \quad [H_i,X^{\pm}_j]=\pm K_{ij}X^{\pm}_j, \quad
[X^{+}_i,X^{-}_j\}={\delta}_{i,j} H_j, \quad (1 \leq i,j \leq 2(n-1)),
\label{aa6}
\end{eqnarray}
where $H_j$ and $X^{\pm}_j$ are the generators of the Cartan subalgebra
and raising/lowering operators, respectively, $K_{ij}$ is the symmetric
Cartan matrix,
\begin{eqnarray}
K_{ij}=K_{ji}= (-1)^{(i{\cal P}+1)}
({\delta}_{i+1,j} - (1+(-1)^{{\cal P}}){\delta}_{i,j}
+(-1)^{{\cal P}}{\delta}_{i,j+1}),
\label{cartan}
\end{eqnarray}
and the brackets $[,\}$ denote the graded commutator.
All its other generators can be derived via the formula
\begin{eqnarray}
Y^{\pm(k+1)}_j=[X^{\pm}_j,[X^{\pm}_{j+1},\ldots[X^{\pm}_{j+k-1},
X^{\pm}_{j+k}\}\ldots \}\}, \quad (1 \leq k < r, \quad 1 \leq j \leq
(r-k)),
\label{alg}
\end{eqnarray}
where $r$ is the rank of the (super)algebra, $r=2(n-1)$.

As opposed to the algebra $sl(2n-1)$, the superalgebra $sl(n|n-1)$
possesses several inequivalent systems of simple roots. The Cartan matrix
\p{cartan} for ${\cal P}=1$ corresponds to a purely fermionic simple--root
system with the fermionic raising/lowering operators $X^{\pm}_j$ which
together with the Cartan generators can be represented by the graded
matrices
\begin{eqnarray}
H_j=(-1)^{j+1}(E_{j,j} + E_{j+1,j+1}), \quad X^{+}_j=(-1)^{j+1}E_{j,j+1},
\quad X^{-}_j=E_{j+1,j}, \quad (E_{i,j})_{pq}
\equiv {\delta}_{i,p}{\delta}_{j,q}
\label{matrreal}
\end{eqnarray}
with the zero (super)trace,
\begin{eqnarray}
str(M) \equiv {\sum}_{j=1}^{2n-1} (-1)^{j+1}M_{jj},
\label{str}
\end{eqnarray}
and the Grassmann parity $d_{ij}$ of their entries defined as
$d_{ij}=(-1)^{i+j}$, where the value $d_{ij}=1$ ($d_{ij}=-1$)
corresponds to bosonic (fermionic) statistics of the entries $M_{ij}$.
At such $Z_2$--grading the generators \p{alg} $Y^{\pm(2k)}_j$ are bosonic
while $Y^{\pm(2k+1)}_j$ are fermionic ones. The Cartan generator of
principal $osp(1|2)$ subalgebra of $sl(n|n-1)$
\begin{eqnarray}
H=\frac{1}{2}{\sum}^{2(n-1)}_{i=1}{\sum}^{2(n-1)}_{j=1}(K^{-1})_{ji}H_{i}
\label{cartan1}
\end{eqnarray}
defines a half-integer grading of superalgebra $sl(n|n-1)=
({\oplus}_{k=1}^{n-1}{\cal G}_{-\frac{k}{2}}){\cal G}_{0}
({\oplus}^{n-1}_{k=1}{\cal G}_{\frac{k}{2}}) \equiv
{\cal G}_{-}{\cal G}_{0}{\cal G}_{+}$, and $H_j \in {\cal G}_{0}$,
$X^{{\pm}}_j \in {\cal G}_{\pm \frac{1}{2}}$ and $Y^{\pm (k)}_j \in
{\cal G}_{\pm \frac{k}{2}}$. Thus, bosonic generators have integer grading,
while fermionic ones have half-integer grading, and positive (negative)
grading corresponds to upper (lower) triangular matrices. Here,
$(K^{-1})_{j,i}$ is the inverse Cartan matrix, $K^{-1}K=KK^{-1}=I$.

The highest weight vector $\ve{j}$ and its dual vector $\vc{j}$
($1\leq j\leq 2(n-1)$) of the $j$--th fundamental representation
possess the following properties:
\begin{eqnarray}
X^{+}_i\ve{j}=0, \quad H_i\ve{j}={\delta}_{i,j}\ve{j},
\quad \vc{j}X^{-}_i=0, \quad \vc{j}H_i={\delta}_{i,j}\vc{j}, \quad
\vc{j}\ve{j}=1.
\label{high}
\end{eqnarray}
The representation is exhibited by repeated applications of the
lowering operators $X^{-}_i$ to $\ve{j}$ and
extraction of all linear--independent vectors with non--zero norm. Its
first few basis vectors are
\begin{equation}
\ve{j}, \quad X^{-}_j\ve{j},  \quad X^{-}_{j\pm 1} X^{-}_j\ve{j}.
\label{vectors}
\end{equation} In the
fundamental representations, matrix elements of the group $G \in
SL(n|n-1)$ ($G \in SL(2n-1)$) satisfy the following important
identity\footnote{Let us remind the definition of the superdeterminant,
$sdet \left(\begin{array}{cc} A, & B \\ C, & D
\end{array}\right) \equiv
det (A-BD^{-1}C ) (det D)^{-1}$.} \cite{ls0}:
\begin{eqnarray}
sdet \left(\begin{array}{cc} \vc{j}X_j^+GX_j^-\ve{j},
& \vc{j}X_j^+G\ve{j} \\
\vc{j}GX_j^-\ve{j}, & \vc{j} G \ve{j} \end{array}\right) =
{\prod}^{2(n-1)}_{i=1}\vc{i} G \ve{i}^{-K_{ji}}, \label{recrel}
\end{eqnarray}
where $K_{ji}$ is the Cartan matrix \p{cartan}. It can be
used as a generating relation for a number of other useful
identities connecting different matrix element of the group $G$.
Indeed, new identities can be derived by replacing $G$ by
$\exp{(t_+l_+)}G\exp{(t_-l_-)}$ on both sides of eqs. \p{recrel}, where
$l_{\pm}$ are arbitrary linear functionals of the generators
$X^{\pm}_i,Y^{\pm(k)}_i$ \p{alg} and differentiating the resulting
expression over the parameters\footnote{The Grassmann parity of $t_{\pm}$
coincides with the parity of the operator $l_{\pm}$.} $t_{\pm}$ at
$t_{\pm}=0$. For example, let us present the following identity:
\begin{eqnarray}
&& sdet \left(\begin{array}{cc} \vc{j}X_j^+l_{+}^pGl_{-}^qX_j^-\ve{j}, &
\vc{j}X_j^+l_{+}^pG\ve{j} \\ \vc{j}Gl_{-}^qX_j^-\ve{j}, & \vc{j} G \ve{j}
\end{array}\right) \nonumber\\
&&= ({\partial}^p_{t_+}{\partial}^q_{t_-}
{\prod}^{2(n-1)}_{i=1}\vc{i} e^{t_+l_+}Ge^{t_-l_-}
\ve{i}^{-K_{ji}})|_{t_{\pm}=0}
\equiv l_{+}^p\circ l_{-}^q\circ
{\prod}^{2(n-1)}_{i=1}\vc{i} G \ve{i}^{-K_{ji}} ~ ~ ~
\label{recrel1}
\end{eqnarray}
which is only valid for the operators $l_{+}$ ($l_{-}$) annihilating the
highest weight vector $\ve{j}$ ($\vc{j}$), $l_{+}\ve{j}=0$
($\vc{j}l_{-}=0$) and $p,q=0,1$. The operation $l_+\circ$ ($l_-\circ$)
is defined by eq. \p{recrel1}, and it represents the left (right)
infinitesimal shift of the group $G$ by the generator $l_{+}$ ($l_{-}$).

The identity \p{recrel} represents a generalization of the famous Jacobi
relation connecting determinants of $(n+1)$, $n$ and $(n-1)$ orders of
some special matrices to the case of arbitrary semisimple Lie
(super)groups. As we will see in the next section, this
identity is so important in deriving integrable mappings and lattices
that one can even say that it is responsible for their existence. We call
it the first Jacobi identity. Besides eqs. \p{recrel}, there is another
independent identity \cite{l}
\begin{eqnarray}
&&(-1)^{{\cal P}}\frac{\vc{j}X_{j}^{+}X_{j-1}^{+}G\ve{j}}{\vc{j} G
\ve{j}}+ \frac{\vc{j-1}X_{j-1}^{+}X_{j}^{+}G\ve{j-1}}{\vc{j-1} G
\ve{j-1}}= \nonumber\\ &&(-1)^{j{\cal P}}
\frac{\vc{j}X_{j}^{+}G\ve{j}\vc{j-1}X_{j-1}^{+}G\ve{j-1}} {\vc{j} G
\ve{j}\vc{j-1} G \ve{j-1}}
\label{recrel3}
\end{eqnarray}
which we use also in what follows and call the second Jacobi identity.
It is responsible for the existence of hierarchies of integrable
equations which are invariant with respect to integrable mappings.
 From this identity one can generate other useful identities in the same way
as it has been explained after formula \p{recrel}.

\section{The SUToda$(2,2;\{s^{+}_j,s^{-}_j\})$ mappings and lattices}
In this section we derive new integrable mappings together with
the corresponding interrupted lattices on the basis of the representation
theory for the algebras $sl(n|n-1)$ (${\cal P}=1$) and $sl(2n-1)$ (${\cal
P}=0$) summarized in the previous section.

Our starting point is the following representation for the group element
$G(z_+,z_-)$ $\in$ \\ $SL(n|n-1)$ ($SL(2n-1)$) \cite{l1,l}:
\begin{eqnarray}
G\equiv M_{+}^{-1}M_{-}
\label{group}
\end{eqnarray}
in terms of the product of upper and lower triangular (including a diagonal)
matrices $M_{+}(z_+)$ and $M_-(z_-)$, respectively, which are defined as
solutions of the following equations:
\begin{eqnarray}
{\cal A}_{\pm}\equiv M_{\pm}^{-1}{\partial}_{\pm} M_{\pm}=
({\mp 1})^{{\cal P}} (\sum_{j=1}^{2(n-1)}({\partial}_{\pm}
{\phi}^{\pm}_j(z_{\pm})H_j+
\nu^{\pm}_{j}(z_{\pm})X^{\pm}_{j}) + \sum_{l=1}^{2(n-1)-1}
s^{\pm}_l(z_{\pm})Y^{\pm (2)}_l),
\label{group1}
\end{eqnarray}
where $z_{\pm}$ are bosonic coordinates
(${\partial}_{\pm} \equiv \frac{\partial}{\partial z_{\pm}}$),
${\phi}^{\pm}_j(z_{\pm})$ and $s^{\pm}_l(z_{\pm})$ are arbitrary
bosonic functions, while $\nu^{\pm}_{j}(z_{\pm})$ are arbitrary fermionic
(bosonic) ones. It is well known that for any finite-dimensional
representations of a (super)algebra, the equations for $M_{\pm}$ can be
integrated in quadratures, and their solutions contain only a finite
number of terms involving products of $X^{\pm}_{j}$ and $Y^{\pm (2)}_l$.
The fields ${\cal A}_{\pm}$ belonging to a (super)algebra can be treated
as components of two different pure gauge connections in the
two--dimensional space with coordinates $(z_+,z_-)$ (see appendix).

At this point it is necessary to make an important, for further
consideration, remark concerning equations \p{group1}. Similar equations
has been used in \cite{l}. They have only but crucial difference
as compared to eqs. \p{group1}: all higher grade functions $s^{\pm}_j$,
which belong to ${\cal A}_{\pm}$, were chosen equal to unity. Of
course, this can always be done by suitable gauge
transformation\footnote{Let us recall that the form of equations
\p{group1} for the matrices $M_{\pm}$ is invariant with respect to gauge
transformations $M_{\pm}\rightarrow g_{\pm}^{-1} M_{\pm} g_{\pm}$
generated by the Cartan subgroup
$g_{\pm}=\exp({\sum}_{j=1}^{2(n-1)}(f^{\pm}_j(z_{\pm})H_j)$, where
$f^{\pm}_j(z_{\pm})$ are parameter--functions.}, but only in the case if
$s^{\pm}_j\neq 0$ for any $j$. Obviously, in the opposite case, i.e. if
$s^{\pm}_l = 0$ for some set of indices $l$, it is impossible to
reach the value $s^{\pm}_l = 1$ for this kind of indices $l$ by any gauge
transformations. Thus, we are led to the conclusion that the space of gauge
inequivalent solutions of eqs. \p{group1} is parameterised by the
numbers $\{s^{+}_j,s^{-}_j, j=1,\ldots,2(n-1)\}$, which take
only two possible values, $0$ or $1$. In general, different sets of
$\{s^{+}_j,s^{-}_j\}$ correspond to inequivalent integrable systems
which can be derived from various fundamental representations of
the group element $G$ \p{group}. Due to this reason as well as to the fact
that UToda$(2,2)$ mapping \cite{l} was derived at the particular choice
$s^{\pm}_j=1$ in eqs.  \p{group1}, it is natural to call them the
SUToda$(2,2;\{s^{+}_j,s^{-}_j\})$ mappings (lattices). In the following
section, we analyze their simplest representatives and demonstrate that
they indeed possess quite different properties and solutions.

Besides the independent functions ${\phi}^{\pm}_j(z_{\pm})$,
$s^{\pm}_l(z_{\pm})$ and $\nu^{\pm}_{j}(z_{\pm})$ entering into equations
\p{group1}, we introduce also a set of dependent fermionic (${\cal P}=1$)
or bosonic (${\cal P}=0$) functions expressed in terms of matrix elements
of fundamental representations of the group $G$ \p{group}:
\begin{eqnarray}
{\mu}^{+}_j(z_{+},z_{-}) \equiv { \vc{j}X_j^+ G\ve{j} \over \vc{j} G \ve{j}},
\quad {\mu}^{-}_j(z_{+},z_{-})\equiv { \vc{j}GX_j^- \ve{j} \over \vc{j} G
\ve{j}}, \quad {\gamma}^{\mp}_{j}(z_{+},z_{-})\equiv
\frac{{\partial}_{\mp}{\mu}^{\pm}_j}{g_j}
\label{def}
\end{eqnarray}
and the bosonic functions\footnote{Hereafter, in order to simplify formulae
we use the following definitions: $X_{-k}^{\pm}=
X_{2(n-1)+k+1}^{\pm}\equiv 0$ ($k\geq 0$), and
$\vc{0}\Gamma\ve{0}=\vc{2(n-1)+1}\Gamma\ve{2(n-1)+1}\equiv 1$
for any group element $\Gamma \in SL(n|n-1)$ ($SL(2n-1)$).}
\begin{eqnarray}
g_j &\equiv& (-1)^{j{\cal P}}{\prod}^{2(n-1)}_{i=1}\vc{i} G
\ve{i}^{-K_{ji}}\nonumber\\
&=&(-1)^{j{\cal P}} \frac{\vc{j-1}G\ve{j-1}^{(-1)^{(j+1){\cal P}}}
\vc{j+1}G\ve{j+1}^{(-1)^{j{\cal P}}}}
{\vc{j} G \ve{j}^{2(1-{\cal P})}}, \quad 1\leq j \leq2(n-1),
\label{def1}
\end{eqnarray}

Our nearest goal is to demonstrate that
the functions $\{g_j,{\gamma}^{+}_{j},{\gamma}^{-}_{j}\}$
satisfy the following closed system of equations:
\begin{eqnarray}
{\partial}_+{\partial}_- \ln g_j &=&
s^{+}_{j+1}s^{-}_{j+1}g_{j+1}g_{j+2}-g_{j}(s^{+}_{j}s^{-}_{j}
g_{j+1}+s^{+}_{j-1}s^{-}_{j-1}g_{j-1})+
s^{+}_{j-2}s^{-}_{j-2}g_{j-1}g_{j-2} \nonumber\\
&+&g_{j+1}{\gamma}^{-}_{j+1}{\gamma}^{+}_{j+1}-
(1+(-1)^{{\cal P}}) g_{j}{\gamma}^{-}_{j}{\gamma}^{+}_{j}+(-1)^{{\cal P}}
g_{j-1}{\gamma}^{-}_{j-1}{\gamma}^{+}_{j-1}, \nonumber\\
{\partial}_{\pm} {\gamma}^{\mp}_{j}&=&s^{\mp}_{j}g_{j+1}
{\gamma}^{\pm}_{j+1}-s^{\mp}_{j-1}g_{j-1}{\gamma}^{\pm}_{j-1}
\label{reseqs}
\end{eqnarray}
with the boundary conditions $g_0=g_{2(n-1)+1}=0$.
They has been called above the \\
SUToda$(2,2;\{s^{+}_j,s^{-}_j\})$ lattice.
The main steps of deriving eqs. \p{reseqs} repeat the corresponding
calculations of Ref. \cite{l} concerning the $sl(n)$--case, but for a
reader's convenience we briefly present them here.

At first, using eqs. \p{high}, identities \p{recrel1} with a group
$G$ \p{group}, and definitions \p{def}--\p{def1}, we obtain other,
equivalent expressions$^{5}$ for ${\gamma}^{\pm}_{j}$,
\begin{eqnarray}
&& {\gamma}^{\mp}_{j}= (-1)^{j{\cal P}}\nu^{\mp}_{j}+
s^{\mp}_{j}{\mu}^{\mp}_{j+1}-s^{\mp}_{j-1}{\mu}^{\mp}_{j-1}, \quad
(0 \leq j \leq 2(n-1)+1),
\label{resinter}
\end{eqnarray}
which being differentiated with respect to $z_{\pm}$ give equations
\p{reseqs} for ${\gamma}^{\pm}_{j}$.

At second, taking into account eqs. \p{high}--\p{vectors} and \p{group}
one can derive the following relation
\begin{eqnarray}
&& {\partial}_+{\partial}_-\ln \vc{j} G \ve{j}=
sdet \left(\begin{array}{cc} \vc{j}X_j^+l_{j+}Gl_{j-}X_j^-\ve{j}, &
\vc{j}X_j^+l_{j+}G\ve{j} \\ \vc{j}Gl_{j-}X_j^-\ve{j}, & \vc{j} G \ve{j}
\end{array}\right)
\label{recrel2}
\end{eqnarray}
with $l_{j\pm}=(\mp)^{{\cal P}+1}
{\nu}^{\pm}_j +s^{\pm}_{j-1} X^{\pm}_{j-1}-(-1)^{{\cal P}}s^{\pm}_{j}
X^{\pm}_{j+1}$. Using identities \p{recrel1}, definitions \p{def}--\p{def1}
and eqs. \p{resinter}, equation \p{recrel2} becomes
\begin{eqnarray}
{\partial}_+{\partial}_- \ln \vc{j} G \ve{j} =(-1)^{(j+1){\cal P}}g_j
(s^{+}_{j}s^{-}_{j}g_{j+1}+ (-1)^{{\cal P}}s^{+}_{j-1}s^{-}_{j-1}
g_{j-1}+{\gamma}^{-}_{j}{\gamma}^{+}_{j}),
\label{eqinter}
\end{eqnarray}
and it can easily be transformed into equation \p{reseqs} for $g_j$.

Thus, we conclude that the SUToda$(2,2;\{s^{+}_j,s^{-}_j\})$
lattice \p{reseqs} is integrable, and its general solutions are given by
formulae \p{group}---\p{def1} and \p{resinter} in terms of matrix elements of
fundamental representations of the (super)groups $SL(n|n-1)$ and $SL(2n-1)$.

To close this section, we would like to stress that the presented here
formalism, underlying the identities between matrix
elements of various fundamental representations of a group, guarantees the
existence of a zero--curvature representation with a pure gauge connection
defined by a single group element $G$ (see appendix). An advantage of the
developed scheme is that starting with a graded algebra and properties of
fundamental representations of the corresponding group we
simultaneously obtain both integrable mappings and general solutions
of their interrupted versions which represent finite-dimensional
exactly integrable lattices.

\section{Examples: supersymmetric mappings}
The consideration of the previous section was based on a pure
super--algebraic level. However, in physical applications integrable systems
possessing supersymmetry cause more interest. In this connection the
important question arises which concerns the existence and
classification of supersymmetric mappings encoded in the above--constructed
super--algebraic integrable systems. We do not know the complete answer to
this question and analyze here only two important representatives of the
SUToda$(2,2;\{s^{+}_j,s^{-}_j\})$ lattices \p{reseqs}, characterized by
\begin{eqnarray}
&& I)  \quad \quad ~
s^{\pm}_j =1, \quad (1 \leq j \leq 2(n-1)), \nonumber\\ && II) \quad \quad
s^{-}_{2j} =s^{-}_{2j+1} =1,
\quad s^{+}_{2j} =1, \quad s^{+}_{2j+1}=0, \quad
(0 \leq j \leq (n-1)).
\label{condn}
\end{eqnarray}
We show that for the $sl(n|n-1)$ superalgebra, i.e. when
${\gamma}^{+}_{j}$ and ${\gamma}^{-}_{j}$ are fermionic fields, they
represent two inequivalent integrable systems which indeed possess higher
supersymmetries, the $N=(2|2)$ and $N=(0|2)$ superconformal symmetries,
respectively. The first system is the $N=(2|2)$ superconformal Toda lattice
equation (see, \cite{eh} and references therein), while the second one
represents two--dimensional generalization of the $1D$ f--Toda chain
\cite{dls,ls2}.

{}~

{\bf 1. The $N=(2|2)$ superconformal Toda lattice.} In the case
of $I)$ \p{condn} equations \p{reseqs} become
\begin{eqnarray}
{\partial}_+{\partial}_- \ln g_j &=&
g_{j+1}g_{j+2}-g_{j}(g_{j+1}+g_{j-1})+g_{j-1}g_{j-2} \nonumber\\
&+&g_{j+1}{\gamma}^{-}_{j+1}{\gamma}^{+}_{j+1}-
(1+(-1)^{{\cal P}}) g_{j}{\gamma}^{-}_{j}{\gamma}^{+}_{j}+(-1)^{{\cal P}}
g_{j-1}{\gamma}^{-}_{j-1}{\gamma}^{+}_{j-1}, \nonumber\\
{\partial}_{\pm} {\gamma}^{\mp}_{j}&=&g_{j+1}{\gamma}^{\pm}_{j+1}-
g_{j-1}{\gamma}^{\pm}_{j-1}, \quad (1 \leq j \leq 2(n-1)).
\label{trlaxeqt2}
\end{eqnarray}

For the $sl(n|n-1)$ superalgebra, i.e. when ${\cal P}=1$ and
${\gamma}^{\pm}_{j}$ are fermionic fields, they coincide with the
component form of the $N=(1|1)$ superconformal Toda lattice equation
\begin{eqnarray}
{\cal D}_{-}{\cal D}_{+} \ln B_j= (-1)^{j+1}(B_{j+1}-B_{j-1})\equiv
{\sum}_{i}K_{ji}B_{i}.
\label{suptod}
\end{eqnarray}
Here, $K_{ji}$ is the symmetric Cartan matrix \p{cartan} of the $sl(n|n-1)$
superalgebra, $B_j(z_+,{\vartheta}_{+};z_-,{\vartheta}_{-})$ is
the bosonic $N=1$ superfield with the components
\begin{eqnarray}
g_{j} \equiv (-1)^{j} B_j|, \quad
{\gamma}^{\pm}_{j} \equiv {\cal D}_{\pm}\ln B_j|,
\label{suptod1}
\end{eqnarray}
where $|$ means the ${\vartheta}_{\pm} \rightarrow 0 $ limit, and
${\cal D}_{\pm}$ are the $N=1$ supersymmetric
fermionic covariant derivatives
\begin{equation}
{\cal D}_{\pm}=\frac{\partial}{\partial {\vartheta}_{\pm}} {\mp}
{\vartheta}_{\pm} {\partial}_{\pm},
\quad {\cal D}_{\pm}^2= {\mp}{\partial}_{\pm}, \quad
\{{\cal D}_+,{\cal D}_-\} =0
\end{equation}
in the $N=(1|1)$ superspace $(z_+, {\vartheta}_+;z_-, {\vartheta}_-)$.

Let us remark that after rescaling $B_j\rightarrow (-1)^{j}B_j$ the factor
$(-1)^{j+1}$ completely disappears from eq. \p{suptod} which
in this case corresponds to anti--symmetric Cartan matrix $K_{ij} =
{\delta}_{i+1,j}-{\delta}_{i,j+1}$. This form of equation \p{suptod} has
been discussed in \cite{ls1} where an infinite family of solutions of its
symmetry equation were constructed. These solutions describe
integrable evolution equations belonging the $N=(1|1)$ superconformal Toda
lattice hierarchy.

The general solutions of the $N=(1|1)$ superconformal Toda
in terms of matrix elements of fundamental
representations of the $SL(n|n-1)$ supergroup with the group element $G$
\p{group}--\p{group1} can easily be derived from formulae
\p{def}---\p{def1} and \p{resinter} and look like
\begin{eqnarray}
&& g_j = (-1)^{j}(\frac{\vc{j+1}G\ve{j+1}}{\vc{j-1}G\ve{j-1}})^{(-1)^{j}},
\nonumber\\
&& {\gamma}^{-}_{j}= (-1)^{j}\nu^{-}_{j}+
{\vc{j+1}GX_{j+1}^-\ve{j+1} \over \vc{j+1} G\ve{j+1}}-
{\vc{j-1}GX_{j-1}^-\ve{j-1} \over \vc{j-1} G\ve{j-1}},  \nonumber\\
&& {\gamma}^{+}_{j}= (-1)^{j}\nu^{+}_{j}+
{ \vc{j+1}X_{j+1}^+ G\ve{j+1} \over \vc{j+1} G\ve{j+1}}-
{ \vc{j-1}X_{j-1}^+ G\ve{j-1} \over \vc{j-1} G\ve{j-1}}.
\label{solutions1}
\end{eqnarray}
These expressions can be promoted to a compact superfield form in terms
of the $N=1$ superfield $B_j(z_+,{\vartheta}_{+};z_-,{\vartheta}_{-})$
\p{suptod}--\p{suptod1},
\begin{eqnarray}
&& ~ ~ ~ ~ ~ ~ ~ ~ ~ ~ ~ ~ ~~ ~ ~ ~ ~ ~ ~ ~ ~ ~ ~
B_j =(\frac{\vc{j+1}{\widetilde G}
\ve{j+1}}{\vc{j-1}{\widetilde G}\ve{j-1}})^{(-1)^{j}}, \nonumber\\
&& {\widetilde G}\equiv
e^{{\vartheta}_{+}\sum_{j=1}^{2(n-1)}({\widetilde \nu}^{+}_jH_j-
(-1)^{j}X^{+}_{j})} Ge^{{\vartheta}_{-}
\sum_{j=1}^{2(n-1)}({\widetilde \nu}^{-}_jH_j-(-1)^{j}X^{-}_{j})},
\quad {\nu}^{\pm}_j\equiv {\widetilde \nu}^{\pm}_{j+1}-
{\widetilde \nu}^{\pm}_{j-1}.
\label{supform}
\end{eqnarray}

Actually, equation \p{suptod} possesses the $N=(2|2)$ superconformal
symmetry and can be rewritten in a manifestly $N=(2|2)$ supersymmetric
form \cite{eh}
\begin{eqnarray}
D_+ D_- \ln \overline \Phi_j=
\Phi_{j-1}-\Phi_{j}, \quad \overline D_+ \overline D_- \ln \Phi_j=
\overline \Phi_{j}- \overline \Phi_{j+1}
\label{suptodn2}
\end{eqnarray}
in terms of two bosonic chiral and antichiral $N=2$ superfields
$\Phi_j(z_+,\theta_+,\overline \theta_+;z_-,\theta_-,\overline \theta_-)$
and $\overline \Phi_j(z_+,\theta_+,\overline \theta_+;z_-,\theta_-,
\overline \theta_-)$, $D_{\pm}\Phi_j=\overline D_{\pm}~ \overline
\Phi_j=0$, respectively, with the components
\begin{eqnarray}
g_{2j} \equiv \overline \Phi_{j}|, \quad
g_{2j+1} \equiv \Phi_{j}|, \quad
{\gamma}^{\pm}_{2j} \equiv D_{\pm}\ln \overline \Phi_{j}|, \quad
{\gamma}^{\pm}_{2j+1} \equiv \overline D_{\pm}\ln \Phi_{j}|.
\label{suptod2}
\end{eqnarray}
Here, $D_{\pm}$ and $\overline D_{\pm}$ are the $N=2$
supersymmetric fermionic covariant derivatives
\begin{eqnarray}
&& D_{\pm}=\frac{\partial}{\partial \theta_{\pm}} {\mp}
\frac{1}{2}\overline \theta_{\pm} {\partial}_{\pm}, \quad
\overline D_{\pm}=\frac{\partial}{\partial \overline \theta_{\pm}}
{\mp}\frac{1}{2} \theta_{\pm} {\partial}_{\pm}, \nonumber\\
&& D_{\pm}^2= {\overline D}_{\pm}^2=\{D_{\pm},\overline D_{\mp}\}=0,
\quad \{D_{\pm},\overline D_{\pm}\} = {\mp}{\partial}_{\pm}
\label{der}
\end{eqnarray}
in the $N=2$ superspace $(z_{\pm}, \theta_{\pm},\overline \theta_{\pm})$.
The $N=2$ superfield solutions can be restored from eqs.
\p{solutions1}--\p{suptodn2} and read as
\begin{eqnarray}
&&\Phi_{j} =\frac{\vc{2j}\Omega\ve{2j}}{\vc{2(j+1)} \Omega\ve{2(j+1)}}, \quad
\overline \Phi_{j} =\frac{\vc{2j+1}\overline \Omega
\ve{2j+1}}{\vc{2j-1}\overline \Omega\ve{2j-1}},
\nonumber\\
&& \Omega\equiv
e^{{\overline \theta}_{+}
\sum_{j=1}^{2(n-1)}({\widetilde \nu}^{+}_jH_j-(-1)^{j}X^{+}_{j})}
G(z_{+}+\frac{1}{2}\theta_{+}\overline \theta_{+},
z_{-}-\frac{1}{2}\theta_{-}\overline \theta_{-})
e^{\overline {\theta}_{-}
\sum_{j=1}^{2(n-1)}({\widetilde \nu}^{-}_jH_j-(-1)^{j}X^{-}_{j})},
\nonumber\\ && \overline \Omega\equiv
e^{{\theta}_{+}
\sum_{j=1}^{2(n-1)}({\widetilde \nu}^{+}_jH_j-(-1)^{j}X^{+}_{j})}
G(z_{+}-\frac{1}{2}\theta_{+}\overline \theta_{+},
z_{-}+\frac{1}{2}\theta_{-}\overline \theta_{-})e^{{\theta}_{-}
\sum_{j=1}^{2(n-1)}({\widetilde \nu}^{-}_jH_j-(-1)^{j}X^{-}_{j})},
\label{supform1}
\end{eqnarray}
where the functions ${\widetilde \nu}^{\pm}_{j}$ are defined in eqs.
\p{supform}. The chiral and antichiral
group elements $\Omega$ and $\overline \Omega$,
$D_{\pm} \Omega=\overline D_{\pm}~ \overline \Omega=0$, are related
by the involution of the algebra of $N=2$ fermionic derivatives \p{der},
\begin{eqnarray}
{\theta}_{\pm}^{*} = {\overline \theta}_{\pm}, \quad
{\overline \theta}_{\pm}~^{*}= {\theta}_{\pm}, \quad
z_{\pm}^{*}=z_{\pm} \quad \Rightarrow \quad
\Omega^{*} = \overline \Omega, \quad
{\overline \Omega}~^{*} = \Omega .
\label{inv}
\end{eqnarray}

For a particular case of the $sl(2|1)$ superalgebra eq. \p{suptodn2}
amounts to the $N=2$ supersymmetric Liouville equation of
Ref. \cite{ik}, where its general solution has been presented. For the
general $sl(n|n-1)$ case the solution of the superconformal Toda lattice
in another parameterisation of the group element and in another basis in
the space of the functions $\{g_j,{\gamma}^{+}_{j},{\gamma}^{-}_{j} \}$
has been found in \cite{as}.

For the $sl(n)$ algebra, i.e. when ${\cal P}=0$ and ${\gamma}^{\pm}_{j}$
are bosonic fields, eqs. \p{trlaxeqt2} reproduce the $UToda(2,2)$ lattice
of ref. \cite{l}.

{}~

{\bf 2. The $N=(0|2)$ superconformal Toda lattice.}
Now, we consider the second system characterized by relation $II)$
\p{condn}. In the new basis  $\{ U_{j}, V_{j}, {\Psi}_{j}, {\overline
{\Psi}}_{j}, {\alpha}_j, {\overline {\alpha}}_{j},
(0 \leq j \leq n-1) \}$
in the space of the fields $\{g_j,{\gamma}^{+}_j, {\gamma}^{-}_j\}$,
\begin{eqnarray}
&& U_j\equiv g_{2j}, \quad \quad \quad \quad {\overline {\Psi}}_{j}
\equiv {\gamma}^{+}_{2j},
\quad \quad \quad \quad {\alpha}_{j}\equiv {\gamma}^{-}_{2j}, \nonumber\\
&& V_{j} \equiv -g_{2j+1}, \quad  {\Psi}_{j} \equiv
-{\gamma}^{+}_{2j+1}, \quad {\overline {\alpha}}_{j} \equiv
-{\gamma}^{-}_{2j+1},
\label{tr4}
\end{eqnarray}
equations \p{reseqs} become
\begin{eqnarray}
&& {\partial}_-{\partial}_+\ln (U_{j}V_{j-1})=
{\partial}_-(\Psi_{j}\overline \Psi_{j}+(-1)^{{\cal P}}
\Psi_{j-1} \overline \Psi_{j-1})-(1+(-1)^{{\cal P}})
(({\partial}_-\Psi_{j})\overline \Psi_{j}+
\Psi_{j-1}{\partial}_- \overline \Psi_{j-1}), \nonumber\\
&& {\partial}_+(\frac{1}{U_j} {\partial}_-\Psi_{j})
=V_{j}\Psi_{j}-V_{j-1}\Psi_{j-1},\quad
{\partial}_+(\frac{1}{V_j} {\partial}_-\overline \Psi_j)=U_{j}\overline
\Psi_{j}- U_{j+1} \overline \Psi_{j+1}, \nonumber\\
&&{\partial}_+{\partial}_- \ln V_j = U_{j}V_{j}-U_{j+1}V_{j+1}+
({\partial}_-\Psi_{j+1}) \overline \Psi_{j+1}-(1+(-1)^{{\cal P}})
\Psi_j{\partial}_-\overline \Psi_j+(-1)^{{\cal P}}
({\partial}_-\Psi_j)\overline \Psi_j, ~ ~ ~ ~ ~
\label{trlaxeqt4}
\end{eqnarray}
\begin{eqnarray}
{\alpha}_j=\frac{1}{U_j}{\partial}_-{\Psi}_j, \quad
{\overline {\alpha}}_j=\frac{1}{V_j}{\partial}_-{\overline {\Psi}}_j
\label{trlaxeqt5}
\end{eqnarray}
with the boundary conditions: $U_0=V_{n-1}=0, \Psi_0=\nu^{+}_{1},
\overline \Psi_{n-1}=\nu^{+}_{2(n-1)}$.

Let us discuss equations \p{trlaxeqt4} in more detail in the case
corresponding the superalgebra $sl(n|n-1)$, i.e. when ${\cal P}=1$ and
fields $\psi_j$ and $\overline \psi_{j}$ are fermionic.

In this case, the first equation of system \p{trlaxeqt4} has the form of a
conservation law with respect to the coordinate $z_{-}$,
\begin{eqnarray}
{\partial}_-{\cal I}=0, \quad {\cal I} \equiv
{\partial}_+\ln (U_{j}V_{j-1})-\Psi_{j}\overline \Psi_{j}+\Psi_{j-1}
\overline \Psi_{j-1},
\label{conserv}
\end{eqnarray}
and as a result, the quantity ${\cal I}$ depends only on the coordinate
$z_{+}$, i.e. ${\cal I}={\cal I}(z_{+})$. It can easily be
expressed in terms of the functions ${\phi}^{+}_j(z_{+})$ and
$\nu^{+}_{j}(z_{+})$, introduced by eqs. \p{group1},
\begin{eqnarray}
{\cal I}(z_+)={\partial}_+\ln
({\eta}_{j-1}(z_+)/{\eta}_{j}(z_+)), \quad {\eta}_j(z_+)
\equiv e^{({\phi}^{+}_{2j}-{\phi}^{+}_{2j+1}
-{\int}{\nu}^{+}_{2j}{\nu}^{+}_{2j+1}dz_{+})},
\label{conserv1}
\end{eqnarray}
where we have used the definitions \p{tr4}, \p{def}--\p{def1},
\p{resinter} and the second Jacobi identity \p{recrel3}.
Keeping in mind relations \p{conserv}--\p{conserv1}, after rescaling
\begin{eqnarray}
U_j \rightarrow u_j \equiv {\eta}_{j} U_j, \quad
V_j \rightarrow v_j \equiv {V_j \over {\eta}_{j}}, \quad
\Psi_j \rightarrow \psi_j\equiv {\eta}_{j} {\Psi_j}, \quad
\overline \Psi_j \rightarrow \overline \psi_j \equiv
{\overline \Psi_j \over {\eta}_{j}}
\label{condtr1}
\end{eqnarray}
we rewrite equations \p{trlaxeqt4} in an equivalent form
\begin{eqnarray}
&& {\partial}_+\ln (u_{j}v_{j-1})=
(\psi_{j}\overline \psi_{j}-\psi_{j-1} \overline \psi_{j-1}),
\nonumber\\
&& {\partial}_+(\frac{1}{u_j} {\partial}_-\psi_{j})
=v_{j}\psi_{j}-v_{j-1}\psi_{j-1},\quad
{\partial}_+(\frac{1}{v_j} {\partial}_-\overline \psi_j)=u_{j}\overline
\psi_{j}- u_{j+1} \overline \psi_{j+1}, \nonumber\\
&& {\partial}_+{\partial}_- \ln v_j=u_{j}v_{j}-u_{j+1}v_{j+1}+
({\partial}_-\psi_{j+1}) \overline \psi_{j+1}-
({\partial}_-\psi_j)\overline \psi_j,
\label{trlaxeq6}
\end{eqnarray}
where the function ${\eta}_j(z_+)$ completely disappears.
These equations reproduce the $1D$ f--Toda equations \cite{dls,ls2} at the
reduction ${\partial}_+={\partial}_-$ to the one--dimensional bosonic
subspace. Therefore, they represent its two--dimensional integrable
generalization. We call them the $2D$ f--Toda lattice. Summarizing all
the above--given formulae we present their general solutions
\begin{eqnarray}
&& u_j={\eta}_{j}{\vc{2j+1}G\ve{2j+1}\over \vc{2j-1}G\ve{2j-1}}, \quad
v_j={1\over {\eta}_{j}}{\vc{2j}G\ve{2j}\over
\vc{2(j+1)}G\ve{2(j+1)}}, \nonumber\\
&& \psi_j= {\eta}_{j}
{\vc{2j}({\nu}^{+}_{2j+1}+X_{2j}^+) G\ve{2j} \over \vc{2j}G\ve{2j}}, \quad
\overline \psi_j \equiv {1\over {\eta}_{j}}
{\vc{2j+1}({\nu}^{+}_{2j} +X_{2j+1}^+) G\ve{2j+1} \over \vc{2j+1}G\ve{2j+1}}
\label{solutions}
\end{eqnarray}
in an explicit form in terms of matrix elements of fundamental
representations of the $SL(n|n-1)$ supergroup with the group element $G$
\p{group}--\p{group1}.

Similar to its one-dimensional counterpart, the 2D f--Toda equations
\p{trlaxeq6} admit the $N=2$ supersymmetry and can be rewritten in the
superfield form
\begin{equation}
F_{j+1} {\overline F_{j+1}}-F_j {\overline F}_j=
{\partial}_+ \ln ((\overline D_- F_{j+1}) (D_- \overline F_j))
\label{ftoda}
\end{equation}
in terms of chiral and antichiral fermionic $N=2$ superfields
$F_j(z_+;z_-, \theta_-,\overline \theta_-)$ and \\
${\overline F}_j(z_+; z_-,\theta_-, \overline \theta_-)$,
$D_-F_j= \overline D_-~ \overline F_j=0$,
respectively, with the components
\begin{eqnarray}
v_j \equiv -D_- {\overline F}_j|, \quad
{\overline {\psi}}_j \equiv  {\overline F}_j|, \quad
u_j  \equiv {\overline D}_-F_j|, \quad
{\psi}_j \equiv  F_j|,
\label{com1}
\end{eqnarray}
where the $N=2$ fermionic derivatives $D_{-}$ and $\overline D_{-}$
are defined in eqs. \p{der}. The general solutions of eq. \p{ftoda} have the
following nice representation$^{5}$:
\begin{eqnarray}
&& F_j= {\eta}_{j}
{\vc{2j}({\nu}^{+}_{2j+1} + X_{2j}^+)
{\cal G}\ve{2j} \over \vc{2j}{\cal G}\ve{2j}}, \quad
\overline F_j \equiv {1\over {\eta}_{j}}{\vc{2j+1}({\nu}^{+}_{2j} +
X_{2j+1}^+) \overline {\cal G}\ve{2j+1} \over \vc{2j+1}
\overline {\cal G}\ve{2j+1}}, \nonumber\\
&& {\cal G} \equiv G(z_{+},z_{-}-\frac{1}{2}\theta_{-}\overline \theta_{-})
e^{-\overline \theta_{-}\sum_{j=1}^{2(n-1)}(-1)^{j}X^{-}_{j}},
\nonumber\\ && \overline {\cal G} \equiv
G(z_{+},z_{-}+\frac{1}{2}\theta_{-}\overline
\theta_{-}) e^{-\theta_{-}\sum_{j=1}^{2(n-1)}(-1)^{j}X^{-}_{j}}.
\label{supsolutions}
\end{eqnarray}
Here, $0\leq j\leq n-1$, ${\cal G}$ and $\overline {\cal G}$ are chiral and
antichiral group elements, $D_{-} {\cal G}= {\overline D}_{-}~
\overline {\cal G}=0$, which are related by involution \p{inv},
${\cal G}^{*} = \overline {\cal G}$, $\overline {\cal G}~^{*}= {\cal G}$.

In the superfield form one can easily recognize that the 2D f--Toda
equations are actually invariant with respect to higher supersymmetry
--- $N=(0,2)$ superconformal symmetry, which generates the transformations
$(z_+;z_-, \theta_-,\overline \theta_-) \rightarrow ({\widetilde
z_+};{\widetilde z_-}, {\widetilde \theta_-}, {\widetilde {\overline
\theta_-}})$,
\begin{eqnarray}
{\widetilde z_+}={\widetilde z_+}(z_+),
\quad \Rightarrow \quad {\partial}_+ = ({\partial}_+ {\widetilde z_+})
{\widetilde {\partial}_+}, \label{vir} \end{eqnarray} \begin{eqnarray} &&
{\widetilde z_-}={\widetilde z_-}(z_-, \theta_-,\overline \theta_-), \quad
{\widetilde \theta_-}={\widetilde \theta_-}(z_-, \theta_-,\overline
\theta_-), \quad {\widetilde {\overline \theta_-}}={\widetilde {\overline
\theta_-}} (z_-, \theta_-,\overline \theta_-), \nonumber\\
&&D_-{\widetilde {\overline \theta_-}}=
{\overline D}_- {\widetilde \theta_-}=0, \quad
D_-{\widetilde z_-}=-\frac{1}{2} {\widetilde {\overline
\theta_-}}D_-{\widetilde \theta_-}, \quad {\overline D}_- {\widetilde
z_-}=-\frac{1}{2} {\widetilde \theta_-}{\overline D}_- {\widetilde
{\overline \theta_-}}, \nonumber\\
&& \Rightarrow D_-=(D_-{\widetilde \theta_-}){\widetilde D}_-, \quad
{\overline D}_-=({\overline D}_-~{\widetilde {\overline \theta_-}})
{\widetilde {\overline D}_-}.
\label{supvir1}
\end{eqnarray}
Under these transformations the superfields $F_j$ and ${\overline F}_j$ are
transforming according to the rule
\begin{eqnarray}
&& F_j(z_+;z_-, \theta_-,\overline \theta_-)
={\varphi}(z_+)({\partial}_+ {\widetilde z_+})^{1-j}
{\widetilde F}_j ({\widetilde z_+};{\widetilde z_-}, {\widetilde
\theta_-}, {\widetilde {\overline \theta_-}}), \nonumber\\
&& {\overline F}_j(z_+;z_-, \theta_-,\overline \theta_-)={\varphi}^{-1}(z_+)
({\partial}_+ {\widetilde z_+})^{j} {\widetilde {\overline F}}_j
({\widetilde z_-};{\widetilde z_-}, {\widetilde \theta_-},
{\widetilde {\overline \theta_-}}),
\label{ftrans}
\end{eqnarray}
where ${\varphi}(z_+)$ is an arbitrary invertible function corresponding
to the local internal $GL(1)$--transformation.
Thus, we conclude that the $2D$ f--Toda lattice \p{ftoda} possesses the
$N=(0|2)$ superconformal symmetry and due to this important property it
can also be called the $N=(0|2)$ superconformal Toda lattice.

Keeping in mind that at ${\cal P}=1$ (i.e. for the case of $sl(n)$ algebra
with bosonic fields $\Psi_j$ and $\overline \Psi_{j}$) equations
\p{trlaxeqt4} represent the integrable bosonic counterpart of the $2D$
f--Toda lattice, it is reasonable to call them the $2D$ bosonic Toda
(b--Toda) lattice equations.

\section{Conclusion}
In the present letter we have demonstrated the dominant role of the
representation theory of graded superalgebras in the context of the problem
of constructing superintegrable mappings. We have shown that the following
chain of relations: representations of a graded algebra $\Rightarrow$
representations of the corresponding group $\Rightarrow$ integrable
mappings (lattices), used in \cite{l} for constructing the bosonic
integrable mappings, perfectly works also in the super--algebraic
case and gives an effective algorithm for constructing new mappings.
We have developed this approach and constructed the new integrable
mappings. All other ingredients of the theory of integrable systems
such as deriving evolution equations, which belong to an integrable
hierarchy, and their multi--soliton solutions are also present in
the framework of this construction. Thus, for example, equations of
hierarchy arise in this approach as solutions of the symmetry
equation which corresponds to an integrable mapping, while their
multi--soliton solutions are related to general solutions of the
corresponding integrable lattices with fixed ends.

We have applied this approach to the case of $sl(n|n-1)$ superalgebra, and
the $2D$ f--Toda lattice possessing the $N=(0|2)$ superconformal symmetry
has been derived for the first time together with its general solutions in
terms of matrix elements of different fundamental representations of the
$SL(n|n-1)$ supergroup.

{}~

\noindent{\bf Acknowledgments.}
A.L. is indebted to the Instituto de Investigaciones en Matem\'aticas
Aplicadas y en Sistemas, UNAM and especially its director Dr. I.Herrera
for beautiful conditions for his work and to S.M.Chumakov and K.B.Wolf
for discussion of the results. A.S. would like to thank L. Bonora and O.
Lechtenfeld for their interest in this paper and useful discussions. This
work was partially supported by the Russian Foundation for Basic Research,
Grant No.  96-02-17634, RFBR-DFG Grant No. 96-02-00180, INTAS Grant No.
93-127 ext..

\section*{Appendix}
\setcounter{equation}{0}
\def\theequation{A.\arabic{equation}}
\vspace{0.5cm}

In this appendix we derive zero--curvature representation for the
SUToda$(2,2;\{s^{+}_j,s^{-}_j\})$ lattice \p{reseqs} (compare with
\cite{l1,ls0}).

Firstly, we rewrite the group element
$G$ defined by eq. \p{group} in the following equivalent form:
\begin{eqnarray}
G\equiv M^{-1}_{+}M_{-}\equiv N_{-}G_0N_{+}.
\label{group2}
\end{eqnarray}
This relation defines the group elements $N_{+}$, $N_{-}$ and $G_0$
representing the exponentiation of the
graded subspaces ${\cal G}_{+}$, ${\cal G}_{-}$ and ${\cal G}_{0}$
of the $sl(n|n-1)$ ($sl(n)$) superalgebra, respectively, in terms of the
group element $G$. It is a simple
exercise to deduce the following two chains of identities:
\begin{eqnarray}
&& {\cal A}_-= G^{-1}{\partial}_{-} G =N_{+}^{-1}
G_0^{-1} (N_{-}^{-1}{\partial}_{-}N_{-}) G_0
N_{+}+N_{+}^{-1}(G_0^{-1}{\partial}_{-}G_0)N_{+}+
N_{+}^{-1}{\partial}_{-}N_{+}, \nonumber\\
&& {\cal A}_+= G{\partial}_{-} G^{-1} =N_{-}
G_0 (N_{+}{\partial}_{-}N_{+}^{-1}) G_0^{-1}
N_{-}^{-1}+N_{-}(G_0{\partial}_{-}G_0^{-1})N_{-}^{-1}+
N_{-}{\partial}_{-}N_{-}^{-1}
\end{eqnarray}
 from relations \p{group2} and \p{group1}.
Comparing the decompositions over graded subspaces of
the right-hand and left-hand sides of
these identities, one can obtain decomposition rules
for the following algebra--valued functions:
\begin{eqnarray}
&& G_0^{-1} (N_{-}^{-1}{\partial}_{-}N_{-}) G_0
\in {\cal G}_{-1} \oplus {\cal G}_{-2}, \quad
G_0^{-1} (N_{-}^{-1}{\partial}_{-}N_{-}) G_0|_{-2} = {\cal A}_-|_{-2},
\nonumber\\ && G_0 (N_{+}{\partial}_{-}N_{+}^{-1}) G_0^{-1}
\in {\cal G}_{+1} \oplus {\cal G}_{+2}, \quad
G_0 (N_{+}{\partial}_{-}N_{+}^{-1}) G_0^{-1}|_{+2} = {\cal A}_+|_{+2},
\label{spectr}
\end{eqnarray}
which we use in what follows. Here, $|_{\pm k}$ means the projection
on the grade subspace ${\cal G}_{\pm k}$.

Secondly, we introduce a new group element ${\cal Q}$,
\begin{eqnarray}
{\cal Q}\equiv
M_{-}N^{-1}_{+}= M_{+}N_{-}G_0,
\label{group3}
\end{eqnarray}
and the components $A_{+}$ and $A_{-}$,
\begin{eqnarray}
&& A_{-}\equiv  {\cal Q}^{-1}{\partial}_{-}{\cal Q}=
G_0^{-1}(N_{-}^{-1}{\partial}_{-}N_{-})G_0 +
G_0^{-1}{\partial}_{-}G_0, \nonumber\\
&& A_{+}\equiv  {\cal Q}^{-1}{\partial}_{+}{\cal Q}=
N_{+}{\partial}_{-}N_{+}^{-1},
\label{zc}
\end{eqnarray}
of a pure gauge by construction connection,
\begin{eqnarray}
[{\partial}_+ -A_{+},{\partial}_- -A_{-}]=0,
\label{lax}
\end{eqnarray}
with the properties
\begin{eqnarray}
&& A_{-} \in {\cal G}_{0}\oplus {\cal G}_{-1} \oplus {\cal G}_{-2}, \quad
A_{-}|_{-2} = {\cal A}_-|_{-2}, \nonumber\\
&& A_{+} \in {\cal G}_{+1} \oplus {\cal G}_{+2}, \quad
A_{+}|_{+2} = G_0^{-1} {\cal A}_+|_{+2} G_0,
\label{spectr1}
\end{eqnarray}
where in deriving these equations, relations \p{group2} and \p{spectr}
have been exploited.

Relations \p{group3}--\p{spectr1} represent the resolved form of the
zero--curvature representation for the SUToda$(2,2;\{s^{+}_j,s^{-}_j\})$
lattice equations \p{reseqs}.

\end{document}